\documentclass[prd,twocolumn,showpacs,preprintnumbers,amsmath,amssymb,floatfix,nofootinbib]{revtex4}

\usepackage{subfigure,graphicx,epsfig,amsmath,amsfonts,amssymb,xcolor,slashed,ulem}

\usepackage{bm}
\usepackage{graphicx}
\usepackage{amsmath}
\usepackage{amsfonts}
\usepackage{amssymb}
\usepackage{color}
\usepackage{ulem}

\usepackage[section]{placeins}

\usepackage{booktabs}
\usepackage[colorlinks, citecolor=blue,anchorcolor=red,menucolor=red, linkcolor=red,filecolor=red,runcolor=red,urlcolor=blue,frenchlinks=red]{hyperref}

\newcommand{\be}{\begin{equation}}
\newcommand{\ee}{\end{equation}}
\newcommand{\ba}{\begin{eqnarray}}
\newcommand{\ea}{\end{eqnarray}}
\newcommand{\nn}{\nonumber}

\newcommand{\mev}{\textrm{ MeV}}
\newcommand{\gev}{\textrm{ GeV}}

\begin{document}
\bibliographystyle{unsrt}
\arraycolsep1.5pt

\title {Discerning the two $K_1(1270)$ poles in $D^0\to \pi^+ V P$ decay}

\author{G. Y. Wang}

\affiliation{Departamento de F\'{\i}sica Te\'orica e IFIC, Centro Mixto
Universidad de Valencia-CSIC, Institutos de Investigacion de
Paterna, Apdo 22085, 46071 Valencia, Spain
}

\affiliation{School of Physics, Zhengzhou University, Zhengzhou, Henan, 450001, China
}

\author{L. Roca}
\affiliation{\it  Departamento de F\'{\i}sica. Universidad de
Murcia. E-30100 Murcia.  Spain.}

\author{E. Oset}

\affiliation{Departamento de F\'{\i}sica Te\'orica e IFIC, Centro Mixto
Universidad de Valencia-CSIC, Institutos de Investigacion de
Paterna, Apdo 22085, 46071 Valencia, Spain}

\begin{abstract}

Within the chiral unitary approach, the axial-vector resonance $K_1(1270)$ has been predicted to manifest a two-pole nature. 
The lowest pole has a mass of $1195\mev$ and a width of $246\mev$ and couples mostly to $K^*\pi$, and the highest pole has a mass of $1284\mev$ and a width of $146\mev$ and couples mostly to $\rho K$. We analyze theoretically how this double-pole structure can show up in the $D^0\to \pi^+ VP$ decays by looking at the vector-pseudoscalar ($VP$) invariant mass distribution for  different $VP$  channels, exploiting the fact that each pole couples differently to different $VP$ pairs.
We find that the final $\bar K^*\pi$ and $\rho \bar K$ channels are sensible to the different poles of the $K_1(1270)$
resonance and hence are suitable reactions to analyze experimentally the double pole nature of this resonance.

\end{abstract}

\maketitle

\section{Introduction}
The standard quark model picture, with baryons of three quarks and mesons made of quark-antiquark, has an undeniable value, putting some order in the large amount of mesons and baryons \cite{isgurmeson,isgurbaryon,capstick,roberts,vijande}. Among the mesons, the vector states follow quite well the $q\bar{q}$ pattern. Recent studies based on QCD and large $N_c$ behaviour give extra support to this picture \cite{sigma}. Yet, many states, as the low lying scalar mesons, many baryons of ${1/2}^-$, ${3/2}^-$ nature and some ${1/2}^+$ excited baryons, do not follow this pattern and call for more complex structures, many of them qualifying as meson-meson or meson-baryon molecules \cite{ulfreview}. Among the meson states, the axial vector meson states are not so successfully reproduced as the vector ones \cite{isgurmeson,vijande}. It is then not so surprising that alternative pictures have emerged, and in \cite{lutz,Roca:2005nm,chengeng} the chiral unitary approach is used, studying the interaction of vector mesons with pseudoscalars, from where the axial vector mesons emerge as a consequence of the interaction, leading to dynamically generated states, or states of vector-pseudoscalar molecular nature. What makes the picture most attractive is that many decays and properties of the axial vector mesons are well reproduced within this picture (see recent works along this line and references therein \cite{liangsakai,dairoca,sakailiang}). 

One of the novel things of the work of \cite{Roca:2005nm} is that two states appeared related to the $K_1(1270)$, which coupled differently to the different channels. These states were further studied  in \cite{Geng:2006yb} in connection with experimental information which provided support for two states, one with $1195\mev$, coupling mostly to $K^*\pi$, and another one with $1284\mev$, coupling mostly to $\rho K$. The two peaks were well differentiated in reactions leading to $K^*\pi$ and $\rho K$ in the final state.

With data piling up on weak meson decays at present hadron facilities, it becomes convenient to exploit the potential of such reactions to provide information on the nature of the axial vector states. Work along these lines for other kind of hadrons in the final state from weak decays of $B$, $D$, $\Lambda_b$, $\Lambda_c$ hadrons has been summarized in \cite{osetrev}. In the present work we pay attention to the $D^0 \to \pi^+ V P$ decay, with $V$ a vector meson and $P$ a pseudoscalar one. We shall see that we can relate the rates of production of $K^*\pi$ and $\rho K$, and the differential mass distributions peak at different masses as a consequence of the different weight of each of the two $K_1(1270)$ resonances in each of the reactions. While present experimental information does not allow a precise determination of the strength of the reactions, the relative rate and the shapes are accurately predicted and the rates are within present experimental reach. We use the results to encourage the experimental community to do a thorough investigation of this problem which can lead to the clear observation of the two $K_1(1270)$ states and important information concerning their nature, and that of the axial vector mesons by extension.

\section{Axial-vector mesons in the chiral unitary approach}
\label{sec:VPunit}

For the sake of completeness, in this section we briefly summarize the  chiral unitary approach for the vector-pseudoscalar interaction  in s-wave \cite{Roca:2005nm,Geng:2006yb}, where most of the low lying axial-vector resonances are obtained  dynamically without the need to include them as explicit degrees of freedom (see refs.~\cite{Roca:2005nm,Geng:2006yb} for further details and explanations).

\begin{figure}[h]
\begin{center}
\includegraphics[width=0.5\textwidth]{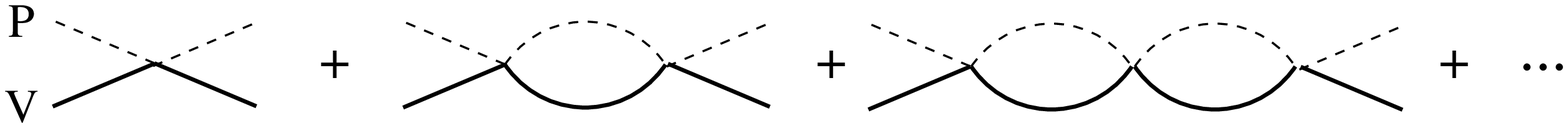}
\caption{\small{Diagrammatic representation 
of the resummation of loops
in the unitarization procedure of the $VP$ interaction.}}
\label{fig:betheVP}
\end{center}
\end{figure}

We make use of the Bethe-Salpeter approach in order to obtain the unitarized $VP$ scattering amplitudes,
\begin{equation}
T=[1+V\hat{G}]^{-1}(-V) \,\vec{\epsilon}\cdot\vec{\epsilon}\,',
\label{bethes}
\end{equation}
which effectively sums the diagrammatic series expressed in
Fig.~\ref{fig:betheVP}.
In Eq.\eqref{bethes}, 
 $\vec{\epsilon}$ and $\vec{\epsilon}\,'$ stand for the vector meson polarizations and
$T$ is a matrix whose element $T_{ij}$ accounts for the scattering amplitude between the $VP$ channels $i$ and $j$, with $i$ and $j$ running for the different $VP$ channels allowed for a given isospin, $I$, strangeness, $S$, and $G$-parity. For the present work we are mostly interested in the 
$I=1/2$, $S=1$, case for which the allowed $VP$ channels are 
$K^* \pi$, $\rho K$, $\omega K$, $\phi K$ and   $K^* \eta$.
But we will also need in the final $VP$ interaction the channels with $I=1$, $S=0$ which can also be classified by its $G$-parity, contributing to the $G=+1$ the channels $\phi\pi$, $\omega \pi$, $\rho\eta$ and $( \bar K^* K +  K^*\bar K)/\sqrt{2}$; and to $G=-1$ the channels $\rho\pi$ and $( \bar K^* K -  K^*\bar K)/\sqrt{2}$.

In Eq.\eqref{bethes}, 
 $\hat{G}=(1+\frac{1}{3}\frac{q^2_l}{M_l^2})G$ is a
diagonal matrix containing the $VP$ loop function for a total incident four-momentum $P$ (equal to  $(\sqrt{s},0,0,0)$ in the center
of mass frame)
\begin{equation}
G_{l}(\sqrt{s})= i \, \int \frac{d^4 q}{(2 \pi)^4} \,
\frac{1}{(P-q)^2 - M_l^2 + i \epsilon} \,
 \frac{1}{q^2 - m^2_l + i
\epsilon}, \label{loop}
\end{equation}
\noindent  Using dimensional regularization, the loop function of Eq.~(\ref{loop}) takes the form
\begin{align}
G_{l}(\sqrt{s})=& \frac{1}{16 \pi^2} \left\{ a(\mu) + \ln
\frac{M_l^2}{\mu^2} + \frac{m_l^2-M_l^2 + s}{2s} \ln
\frac{m_l^2}{M_l^2} \right. \nonumber\\ 
&+ \frac{q_l}{\sqrt{s}} \left[ \ln(s-(M_l^2-m_l^2)+2
q_l\sqrt{s}) \right. \nn\\
&+
\ln(s+(M_l^2-m_l^2)+2 q_l\sqrt{s})  \nn  \\
&  
 - \ln(-s+(M_l^2-m_l^2)+2 q_l\sqrt{s}) \nn\\
&\left.\left.- \ln(-s-(M_l^2-m_l^2)+2 q_l\sqrt{s}) \right]\right\},
\label{propdr}
\end{align}
\noindent
where $M_l$($m_l$) is the mass of the vector(pseudoscalar) meson,
$q_l$ is the on-shell momentum of the meson in the loop,
$\mu$ is the scale of dimensional regularization and $a(\mu)$  is a subtraction constant which reabsorbs possible changes in $\mu$ so
that the model only depends on one free parameter, $a(\mu)$.
By fitting to experimental $K^-
p\rightarrow K^-\pi^+\pi^- p$ data, in Ref.~\cite{Geng:2006yb} the value $a=-1.85$, for $\mu=900\mev$ was obtained, which is indeed of natural size \cite{Roca:2005nm}.
The effect of the finite width of the vector mesons can be taken into account by folding the loop functions with the vector-meson spectral function
\begin{equation}
S(s_V)=-\frac{1}{\pi}\mathrm{Im}\left\{\frac{1}{s_V-M^2_V+iM_V\Gamma_V}\right\}
\end{equation}
such that
\begin{align}
G_{l}(\sqrt{s},M_l,m_l)=\frac{\int\limits^{(M_V+2\Gamma_V)^2}_{(M_V-2\Gamma_V)^2}ds_V
G(\sqrt{s},\sqrt{s_V},m_l) S(s_V)}{\int\limits^{(M_V+2\Gamma_V)^2}_{(M_V-2\Gamma_V)^2}ds_V
S(s_V)}
\label{eq15}
\end{align}
where we have taken a reasonable range, $(M_V\pm 2\Gamma_V)^2$, in the $s_V$ integration.

The other input needed in Eq.\eqref{bethes} 
is the $VP\to VP$ tree level amplitudes, $V_{ij}$, which can be obtained from the chiral invariant Lagrangian \cite{Birse:1996hd,Roca:2005nm}
\begin{equation}
{\cal L}_{VVPP}=-\frac{1}{4f^2}
\mathrm{Tr}\left([V^{\mu},\partial^{\nu}V_{\mu}]
        [P,\partial_{\nu}P]\right),
\label{eq:L}
\end{equation}
\noindent where $V$ and $P$ are the usual $SU(3)$ matrices containing vector and pseudoscalar mesons respectively.

The s-wave projected $VP\to VP$ tree level amplitudes obtained from  the Lagrangian of Eq.~\eqref{eq:L} is 
\begin{eqnarray}
V_{ij}(s)&=&-\frac{\epsilon\cdot\epsilon'}{8f^2} C_{ij}
\left[3s-(M_i^2+m_i^2+M_j^2+m_j^2)\right.\nonumber\\
&&\left.\hspace{1.5cm}-\frac{1}{s}(M_i^2-m_i^2)(M_j^2-m_j^2)\right],
\label{eq:Vtree}
\end{eqnarray}
with $f=115\mev$ \cite{Geng:2006yb} and 
$C_{ij}$ are numerical coefficients which are tabulated in \cite{Roca:2005nm,Geng:2006yb}. The function $V$ entering Eq.~\eqref{bethes} is given by Eq.~\eqref{eq:Vtree} removing the $\epsilon\cdot\epsilon'$ factor.

Note that we have not included explicitly axial-vector resonances  in the formalism, (by means of Breit-Wigner amplitudes
 or any other approaches). The axial-vector resonances are actually generated dynamically from the highly non-linear dynamics involved in the unitary amplitude, Eq.~\eqref{bethes},  where the only input is the lowest order $VP$ chiral Lagrangian, Eq.~\eqref{eq:L}.
Indeed, they show up as poles of the scattering amplitude, Eq.~\eqref{bethes}, in its second Riemann sheet of the complex center of mass energy, $\sqrt{s}$, plane.
For the $(S,I)=(1,1/2)$ case, two poles are found at $(1195-i123)\mev$ and $(1284-i73)\mev$, where the real part can be identified with the mass and the imaginary part with half the width. In \cite{Roca:2005nm,Geng:2006yb} these poles were assigned to the experimental $K_1(1270)$ resonance, which therefore would actually correspond to two different resonances.
For the $(S,I)=(0,1)$ with negative $G$-parity one pole was found in \cite{Roca:2005nm} associated to the $a_1(1260)$ and another one with negative $G$-parity associated to the $b_1(1235)$.

The couplings, $g_i$, of the resonances associated to the poles to the $i$-th $VP$ channel can be obtained from  the residue of the $T_{ij}$ amplitude at the pole position, $\sqrt{s_p}$, since close to the pole the amplitude $T_{ij}$ in the second Riemann sheet takes the form
\begin{equation}
T_{ij}=\frac{g_i g_j}{s-s_p}.
\label{eq:tijpoles}
\end{equation}
The couplings obtained for $(S,I)=(1,1/2)$  are shown in Table~\ref{tab:coup}.
\begin{table}[h]
\caption{\small{Couplings of the two poles associated to the $K_1(1270)$ resonance  to the different $VP$ channels. All the units are in MeV.}}
\begin{center}
\begin{tabular}{|c|cc|cc|}
\hline &&&& \\[-4.5mm]
 $\sqrt{s_p}$ & \multicolumn{2}{c|}{$1195-i123$} &\multicolumn{2}{c|}{ $1284-i73$ } \\
\cline{2-5}
& $g_i$ & $|g_i|$ & $g_i$ & $|g_i|$ \\
\hline  &&&& \\[-4.5mm]
$\phi K$   &  $2096-i1208$  & $2420$ & $1166-i774$  & $1399$  \\
$\omega K$ &  $-2046+i821$ & $2205$ & $-1051+i620$ & $1220$  \\
$\rho K$   &  $-1671+i1599$& $2313$ & $4804+i395$  & $4821$  \\
$K^* \eta$ &  $72+i197$    & $210$  & $3486-i536$   & $3526$  \\
$K^* \pi$  &  $4747-i2874$ & $5550$ & $769-i1171$   & $1401$  \\
\hline
\end{tabular}
\label{tab:coup}
\end{center}
\end{table}
We can see that the lowest mass pole couples mostly to $K^*\pi$ and the highest mass pole couples dominantly to $\rho K$, although the couplings to most of the other channels are also sizable. Therefore we would expect that different reactions weighing differently the $K^*\pi$
and $\rho K$ production mechanisms would see different shapes, corresponding to one or the other pole associated to the $K_1(1270)$ resonance
 \cite{Roca:2005nm}.

\begin{figure}[h]
     \centering
     \subfigure[]{
          \includegraphics[width=.95\linewidth]{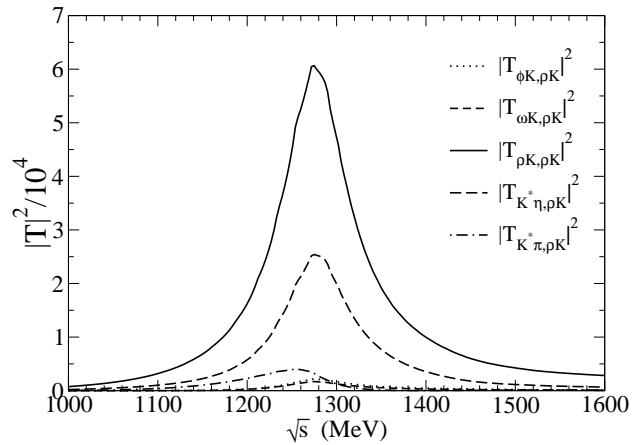}}\\
     \subfigure[]{
          \includegraphics[width=.95\linewidth]{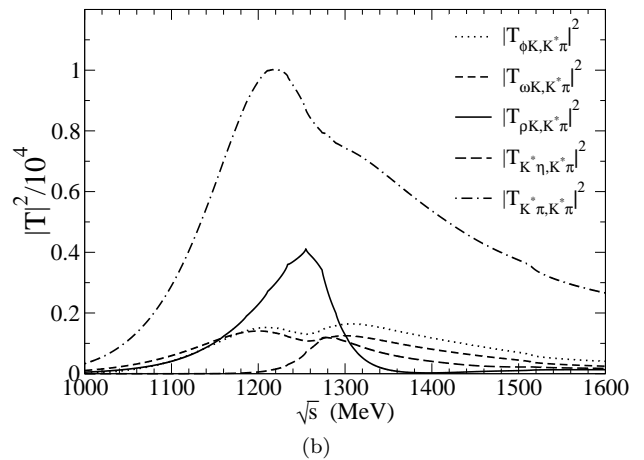}}
    \caption{\small{Modulus squared of the $VP$ scattering amplitudes for $(S,I)=(1,1/2)$} }\label{fig:TVP}
\end{figure}
Indeed, in Fig.~\ref{fig:TVP}  we show the amplitudes in the real axis, for $(S,I)=(1,1/2)$, of the different $VP$ channels going to  $\rho K$ and $K^*\pi$. The effect of the different poles in the shape of the mass distributions for the different channels is clearly visible. 
For instance, in the $\rho K\rightarrow \rho K$ amplitude we can recognize the dominance of the higher mass pole and in  
the $K^*\pi\rightarrow K^*\pi$ channel the lower mass one.
This dependence on the final channels of the weight of the different poles is what we expect to observe in  the present work in the $VP$ mass distributions in $D^0\to \pi^+\rho \bar K$ and $D^0\to \pi^+\bar K^*\pi$ decays.
For $(S,I)=(0,1)$ channels, the values of the pole positions, the couplings to the different $VP$ channels and plots for the $VP$  amplitudes can be see in ref.~\cite{Roca:2005nm}

\section{Formalism for $D^0\to \pi^+ VP$ decay}

\subsection{Tree level production}

\begin{figure}[h]
\begin{center}
\includegraphics[width=0.5\textwidth]{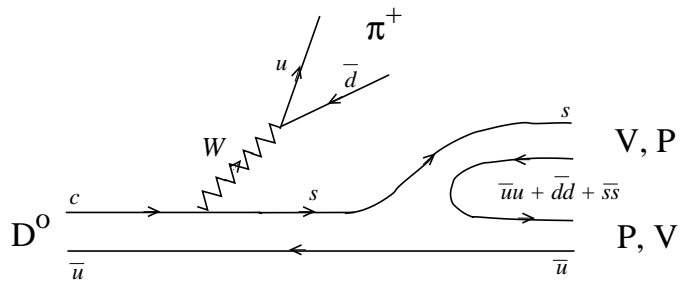}
\caption{\small{Elementary $D^0\to \pi^+ VP$ process at the quark level}}
\label{fig:Tquarks}
\end{center}
\end{figure}

The dominant contribution to the elementary $D^0\to \pi^+ VP$ process at the quark level is depicted in Fig.~\ref{fig:Tquarks}.  First the charm quark of the $D^0$ meson produces an $s$ quark  and a $\pi^+$ through the Cabibbo dominant vertices $Wcs$ and $Wu\bar d$. Production of a kaon instead of the pion would imply a Cabibbo suppressed coupling to the $W$ boson.
The hadronization, giving rise to a $VP$ final pair, is then implemented with  the $^3P_0$ model  \cite{micu,LeYaouanc:1972vsx,bijker}, where an extra $\bar{q} q$ with the quantum numbers of the vacuum is produced and becomes real thanks to  the
 large phase space available.
For the different $VP$  pairs that can be produced, we can determine their relative strength
 by using the following $SU(3)$ argument:

After the
hadronization, the quark flavor final state is given by
\begin{align*}
|H\rangle \equiv |s\,(\bar u u +\bar d d +\bar s s)\,\bar u \rangle.
\end{align*}
If we define
\begin{equation}
q\equiv \left(\begin{array}{c}u\\d\\s\end{array}\right)\,\text{~~and~~~}
M\equiv q\bar q^\intercal=\left(\begin{array}{ccc}u\bar u & u\bar d & u\bar s\\
					       d\bar u & d\bar d & d\bar s\\
					       s\bar u & s\bar d & s\bar s
\end{array}\right)\,,
\label{eq:Mqqbar}
\end{equation} 
then the final hadron state can be written as
\begin{align}
|H\rangle = \sum_{i=1}^3{|M_{3i}q_i\bar u} \rangle=
            \sum_{i=1}^3{|M_{3i}M_{i1} }    \rangle=
	     |(M^2)_{31}\rangle .
\end{align}

 Next we write the $q\bar q$ $M$-matrix in terms of the physical mesons $P$, or $V$, as
\begin{align*}
M\Rightarrow P=
\left(\begin{array}{ccc} 
              \frac{\pi^0}{\sqrt{2}}  + \frac{\eta}{\sqrt{3}}+\frac{\eta'}{\sqrt{6}}& \pi^+ & K^+\\
              \pi^-& -\frac{1}{\sqrt{2}}\pi^0 + \frac{\eta}{\sqrt{3}}+ \frac{\eta'}{\sqrt{6}}& K^0\\
              K^-& \bar{K}^0 & -\frac{\eta}{\sqrt{3}}+ \frac{2\eta'}{\sqrt{6}} 
      \end{array}
\right)\,,
\label{eq:Pmatrix}
\end{align*}
or the vector-mesons
\begin{equation}
 M\Rightarrow V \equiv
 \left(\begin{array}{ccc} \frac{1}{\sqrt{2}} \rho^0 +
\frac{1}{\sqrt{2}}\omega
 & \rho^+ & K^{*+}\\
\rho^-& - \frac{1}{\sqrt{2}} \rho^0 + \frac{1}{\sqrt{2}}\omega
& K^{*0}\\
K^{*-}& \bar{K}^{*0} & \phi
\end{array}
\right) . \label{eq:Vmatrix}
\end{equation}
In $P$ the usual mixing between the singlet and octet to give $\eta$ and $\eta'$ \cite{Bramon:1992kr}, and for $V$ the
ideal $\omega_1$-$\omega_8$ mixing to produce $\omega$ and $\phi$, have been used in order to agree with the quark content of $M$ in Eq.~(\ref{eq:Mqqbar}).
Thus, in terms of $V$ and $P$ fields, for the final hadronic state we find
\begin{align}
(VP)_{31}+(PV)_{31}=&\frac{1}{\sqrt{2}} {K^*}^-\pi^0+\bar{K}^{*0} \pi^-+\phi K^- \nn\\
+&\frac{1}{\sqrt{2}}\rho^0 K^-+\frac{1}{\sqrt{2}}\omega K^-+\bar {K}^0\rho^-,
\label{eq:weightsVP}
\end{align}
where, as in \cite{Roca:2005nm}, we neglect the $\eta'$ because of its large mass. Eq.~\eqref{eq:weightsVP} provides the relative weights between the different $VP$ channels in charge basis, and  can be converted to $VP$ states in isospin basis with $I=1/2$, $I_3=-1/2$, which is the basis used in the previous section, giving rise to:
\begin{align}
&\sum_i h_i\, (V_iP_i)\Big|_{\left(I=\frac{1}{2},I_3=-\frac{1}{2}\right)}\equiv \nn \\
&\sqrt{\frac{3}{2}} \bar K^*\pi-\sqrt{\frac{3}{2}} \rho \bar K +\frac{1}{\sqrt{2}} \omega \bar K + \phi \bar K.
\label{eq:defhi}
\end{align}

Taking into account the previous discussion, the tree level amplitude for the $D^0\to \pi^+ V_iP_i$ decay can then be written as 
\begin{align}
t_i\equiv A h_i,
\end{align}
with $A$ a normalization constant.
Note, however, that since we have a vector meson in the final state the amplitude must also be proportional to its polarization vector $ \epsilon$, which must be contracted with another vector.
The other vector must be a momentum of the particles. Taking into account that  $\epsilon_\mu p_V^\mu=0$ from the Lorentz condition and that $p_{D^0}=p_{\pi^+}+p_V+p_P$, we are only left with two independent structures:  $\epsilon_\mu p_{\pi^+}^\mu$ and  $\epsilon_\mu p_P^\mu$. In addition, we can safely ignore the
$\epsilon^0$ component of the vector polarization, which is found to be a very good approximation in these kind of processes as shown in  \cite{Sakai:2017hpg} (see Appendix A of this reference).
Thus the tree level amplitude will have the structure
\begin{align}
t_{\textrm{tree}}= a h_i \vec \epsilon\cdot\vec p_{\pi^+}
                   +b h_i\vec \epsilon\cdot\vec p_P.
\label{eq:aplusb}		   
\end{align}		   
with $a$ and $b$ unknown complex constants which will be determined later on.

\subsection{$VP$ final state interaction}

\begin{figure}[h]
\begin{center}
\includegraphics[width=0.45\textwidth]{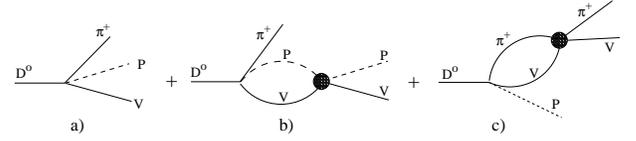}
\caption{\small{$VP$ final state interaction.}}
\label{fig:TFSI}
\end{center}
\end{figure}

As explained in section~\ref{sec:VPunit}, the two poles for the axial-vector $K_1(1270)$ resonance are dynamically generated from the $VP$ interaction. Therefore we have to implement the final state interaction of the $VP$ pair produced in the elementary $D^0\to \pi^+ V P$ mechanism of Fig.~\ref{fig:Tquarks}. This is depicted in Fig.~\ref{fig:TFSI}, where the thick dot represents the $VP$ scattering amplitude accounting for the series  of Fig.~\ref{fig:betheVP}.
Actually we will see below that the b) mechanism will generate the two $K_1(1270)$ poles and the c) mechanism will only be possible for the $I=1$ channels that generate the $a_1(1260)$ and $b_1(1235)$. Furthermore, note that the tree level a) mechanism contributes to the $a$ and $b$ terms of Eq.~\eqref{eq:aplusb}, the b) mechanism only to the $a$ term and the c) mechanism only to the $b$ term.

The $D^0\to \pi^+ V P$ channels for which the $K_1(1270)$ is more relevant are those  where the allowed $VP$ invariant mass distribution contains or is close to the region of the poles associated to this resonance. The thresholds for the $VP$ channels in  $I=1/2$, $S=1$, $K^* \pi$, $\rho K$, $\omega K$, $\phi K$ and   $K^* \eta$ are $1032\mev$, $1271\mev$, $1278\mev$, $1515\mev$ and $1441\mev$ respectively. Therefore $K^* \pi$, $\rho K$ and $\omega K$ are the final channels which contain or are closer to the real part of the two $K_1(1270)$ poles, ($1195\mev$ and $1284\mev$), and then we expect the $K_1(1270)$ to play an important role in those channels. 
Thus, in the present work we are going to evaluate the channels
$D^0\to \pi^+\bar{K}^{*0}\pi^-$,
$D^0\to \pi^+\rho^0 K^-$ and $D^0\to \pi^+ \omega K^-$.

The scattering amplitudes, ${\cal M}_i$, for the $D^0\to \pi^+ V_iP_i$ for these process can be written as

\begin{align}
{\cal M}_{D^0\to \pi^+\bar{K}^{*0}\pi^-}&=a \vec \epsilon\cdot\vec p_{\pi^+} \sqrt{\frac{2}{3}}\left(h_{\bar {K^*\pi}}+\sum_j h_j G_j T^{\rm{I=1/2}}_{j,K^*\pi}\right)\nn \\
&+ b \vec \epsilon\cdot\vec p_{\pi^-} \sqrt{\frac{2}{3}}h_{\bar K^*\pi},\nn\\
{\cal M}_{D^0\to \pi^+\rho^0 K^-}&=a \vec \epsilon\cdot\vec p_{\pi^+} \left(\frac{-1}{\sqrt{3}}\right)\left(h_{\rho \bar K}+\sum_j h_j G_j T^{\rm{I=1/2}}_{j,\rho \bar K}\right)\nn \\
&+ b \vec \epsilon\cdot\vec p_{K^-}\left(\frac{-1}{\sqrt{3}}\right)h_{\rho \bar K}\left(1+\frac{1}{2}G_{\rho \pi} T^{\rm{I=1}}_{\rho\pi,\rho \pi}\right),\nn\\
{\cal M}_{D^0\to \pi^+\omega K^-}&=a \vec \epsilon\cdot\vec p_{\pi^+} \left(h_{\omega \bar K}+ \sum_j h_j G_j  T^{\rm{I=1/2}}_{j,\omega \bar K}\right)\nn \\
&+ b \vec \epsilon\cdot\vec p_{K^-}(h_{\omega \bar K}  +h_{\omega \bar K}G_{\omega \pi} T^{\rm{I=1}}_{\omega\pi,\omega \pi}\nn\\
&+h_{\phi K} G_{\phi \pi} T^{\rm{I=1}}_{\phi\pi,\omega \pi}),
\label{eq:MTunit}
\end{align}
with $h_i$ the coefficients defined in Eq.~\eqref{eq:defhi}, $G_i$ the same $VP$ loop function of
Eq.~\eqref{eq15} and the index $j$ runs for all possible different $VP$ channels. The superindex in the $VP$ amplitudes, $T_{ij}$ stand for the total isospin. Again, a folding with the vector meson spectral function is implemented for  $G_j$.

For the $VP$ unitarized amplitudes $T_{ji}$ one should in principle use the amplitudes obtained from Eq.\eqref{bethes}, (see again Fig.~\ref{fig:TVP}). But in an actual experimental analysis one typically would try to fit Breit-Wigner like shapes for the resonance and then one would use something more similar to 
Eq.\eqref{eq:tijpoles}.
Therefore, in Eq.~\eqref{eq:MTunit}, we can also approximate the $T_{ji}$ amplitudes by Eq.\eqref{eq:tijpoles} and then use, for $I=1/2$, $S=1/2$, 
 \begin{align}
T_{ji}=T^A_{ji}+T^{B}_{ji}
\label{eq:MTBW}
\end{align}
where the superindex $A$ stands for the amplitude of Eq.\eqref{eq:tijpoles} for the lowest mass $K_1(1270)$ pole and  the superindex $B$ for the highest mass pole. Similarly for the 
$I=1$, $S=0$, amplitudes we can use Breit-Wigner shapes as in Eq.\eqref{eq:tijpoles}. Note that the closer the poles are to the real axis the better this approximation is, and it is exact 
 at the poles.
 For $I=1$, where the unitarized amplitudes  generate dynamically the $a_1(1260)$ and $b_1(1235)$ resonances, we use instead the Breit-Wigner shapes with couplings from \cite{Roca:2005nm} and masses and widths from the PDG \cite{pdg}.

The amplitudes  of Eq.~\eqref{eq:MTunit} are of the form
\begin{align}
{\cal M}_i= A \vec \epsilon\cdot\vec p_{\pi^+}
                   +B\vec \epsilon\cdot\vec p_P.
\label{eq:Maplusb}		   
\end{align}	
For the evaluation of the  $D^0\to \pi^+ V_iP_i$ decay width we will have to 
perform
the sum over polarizations of the modulus squared, $\sum|{\cal M}_i|^2$, which gives
\begin{align}
\sum|{\cal M}_i|^2=|A|^2 \vec p\,^2_{\pi^+} +|B|^2 \vec p\,^2_P + 2 \textrm{Re}(A B^*)
\, \vec p_{\pi^+}\cdot\vec p_P.
\label{eq:tiw}
\end{align}
with 
$\vec p_{\pi^+}$ and $\vec p_P$ the momentum of the $\pi^+$ and the pseudoscalar $P_i$ in the $D_0$ rest frame.
From Eq.~\eqref{eq:tiw}, the invariant mass distributions $\frac{d \Gamma(D^0\to \pi^+ V_i P_i)}{d M_{V_iP_i}}$ and $\frac{d \Gamma(D^0\to \pi^+ V_i P_i)}{d M_{V_i\pi^+}}$ are readily obtained after integrating over one invariant mass the double differential mass distribution \cite{pdg}
\begin{align}
\frac{d^2 \Gamma(D^0\to \pi^+ V_i P_i)}{d M_{V_iP_i} d M_{V_i\pi^+}}= 
\frac{1}{8(2\pi)^3 M_{D^0}^3}\sum|{\cal M}_i|^2 M_{V_iP_i}M_{V_i\pi^+}.
\end{align}
In order to take into account the finite width of the final vector meson, a folding with the vector meson spectral function, analogous to the one in Eq.~\eqref{eq15}, is implemented for the invariant mass distributions. Indeed, for the $\rho$ vector meson, the threshold of the $\rho \bar K$ mass distribution is very close to the position of the two $K_1(1270)$ poles and then the tail of the $\rho$ meson is very relevant.

The amplitudes in Eq.~\eqref{eq:MTunit} depend on the complex $a$ and $b$ constants. We can take $a$ real because of an unobservable global phase, and then we are left with three unknown parameters: $a$, $|b|$ and $\phi_b$, with $\phi_b$ the phase of $b$.
We could in principle determine these parameters from experimental $D^0\to \pi^+ V_i P_i$ branching ratios data in the PDG \cite{pdg}.
However, we cannot use experimental data for $BR(D^0\to \pi^+ V_i P_i)$ coming from an intermediate $K_1(1270)$ since it is incompatible  with our model because the experimental analyses are performed assuming that there is only one $K_1(1270)$ pole instead of the two obtained in our theory.
 Therefore we are only left with the following available experimental information in the PDG \cite{pdg}: First, the branching ratio 
$BR(D^0\to \pi^+\rho^0 K^-)=(6.87\pm 0.31)\%$  to which we have to subtract 
$BR(D^0\to  \bar{K}^{*0}  \rho^0; \bar{K}^{*0}\to K^-\pi^+)=(1.01\pm0.05)\%$
 since that channel does not contribute to the mechanism of Fig.~\ref{fig:Tquarks}.
We will call that subtracted value $BR_1\equiv\widetilde{BR}(D^0\to \pi^+\rho^0 K^-)=(5.86\pm0.31)\%$. The other experimental datum we can use from the PDG \cite{pdg} is 
$BR(D^0\to \pi^+ \omega K^-)=(3.1\pm0.6)\%$, to which we have to subtract $\frac{2}{3}BR(D^0\to  \bar{K}^{*0}  \omega)=\frac{2}{3}(1.1\pm0.5)\%$ which we will call  $BR_2\equiv\widetilde{BR}(D^0\to\pi^+ \omega K^-)=2.4\pm0.7\%$. With two data and three parameters there could be mathematically infinitely many solutions, but that would not be a problem if they turned out to lay on a  narrow range of values. Actually, in the present case for the concrete data explained above, we do not find exact mathematical solution for the central values of the experimental $BR_1$ and $BR_2$ but there is solution within the uncertainty range of the experimental data. In order to get this solution we minimize the chi-squared function 
\begin{align}
\chi^2= \left(\frac{BR_1^{\textrm{exp}}-BR_1^{\textrm{th}}}
{\sigma_{BR_1}}\right)^2+
\left(\frac{BR_2^{\textrm{exp}}-BR_2^{\textrm{th}}}
{\sigma_{BR_2}}\right)^2
\label{eq:chi2}
\end{align}
where the labels ``exp" and ``th" stand for the experimental and theoretical values and $\sigma_{BR_i}$ are the experimental uncertainties mentioned above.
We get a minimum value of $\chi^2=0.26$ at $a=3.13\gev^{-1}$, $b=5.47e^{ 1.74\pi i}\gev^{-1}$ for which $BR_1=5.82\%$ and $BR_2=2.75\%$ well within the range allowed by the uncertainties of the experimental values. In any case, note that in the present work we are specially interested in the shape of the invariant mass distribution to see the different shape of the two distinct poles of the $K_1(1270)$ and the global normalization is a collateral aspect.

\section{Results}

\begin{figure*}[h]
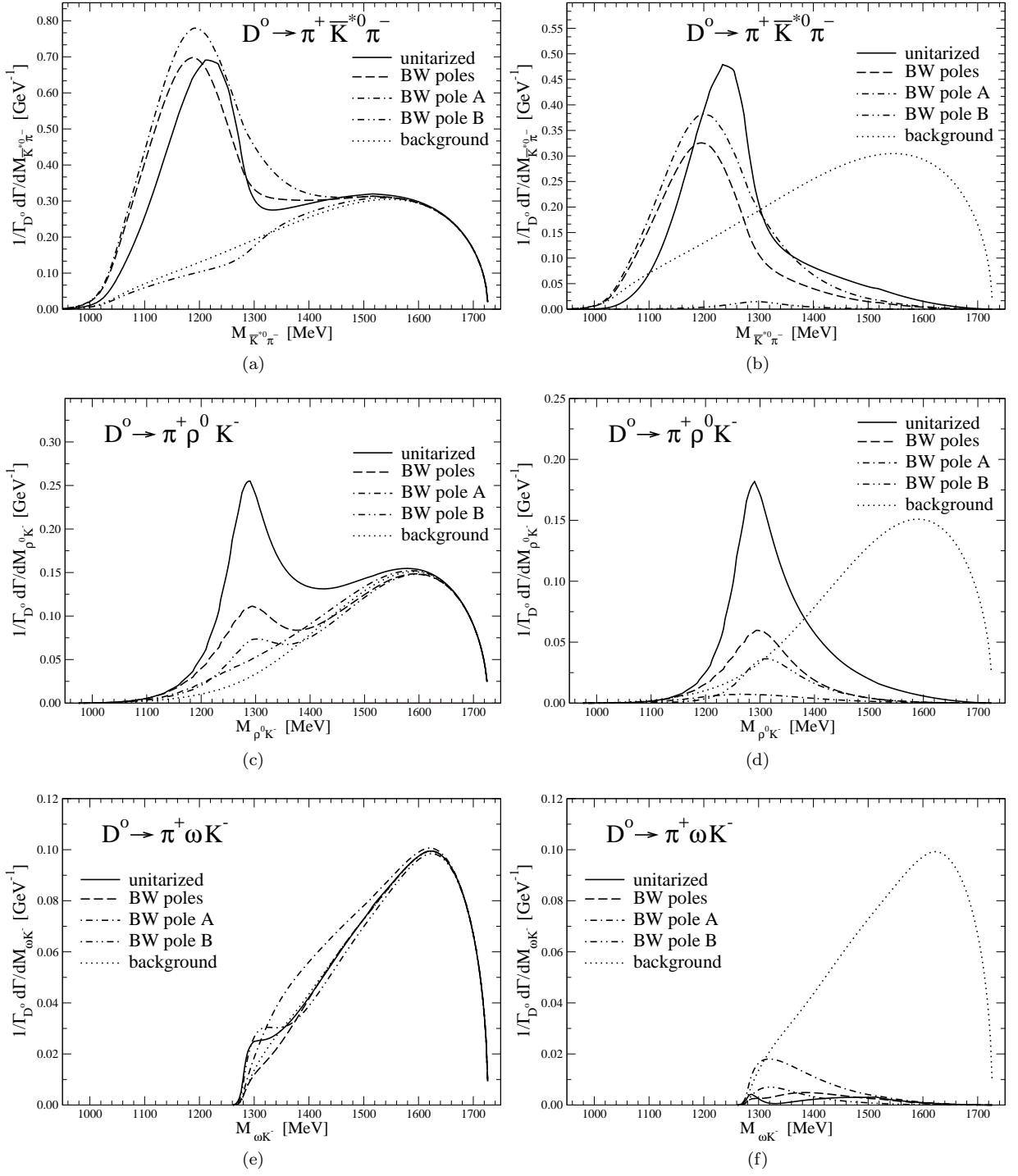

     \centering
     \subfigure[]{\includegraphics[width=.45\linewidth]{M1a.eps}}
     \subfigure[]{\includegraphics[width=.45\linewidth]{M1b.eps}} \\
     \subfigure[]{\includegraphics[width=.45\linewidth]{M2a.eps}}
     \subfigure[]{\includegraphics[width=.45\linewidth]{M2b.eps}} \\
     \subfigure[]{\includegraphics[width=.45\linewidth]{M3a.eps}}
     \subfigure[]{\includegraphics[width=.45\linewidth]{M3b.eps}} \\
   \caption{\small{VP invariant mass distribution in the $D^0\to \pi^+ VP$ decay for $D^0\to \pi^+\bar{K}^{*0}\pi^-$,
$D^0\to \pi^+\rho^0 K^-$ and $D^0\to \pi^+ \omega K^-$. Left panels: considering the interaction with the tree level elementary mechanism of Fig.~\ref{fig:Tquarks}. Right panels: without the interference with the background contributions. Further explanations in the text.
}}
\label{fig:results1}
\end{figure*}

In Fig.~\ref{fig:results1} left we show the $VP$ invariant mass distribution for the $\bar{K}^{*0}\pi^-$, $\rho^0 K^-$ and $\omega K^-$ final $VP$ channels considering the contribution of different  mechanisms. The different curves are explained and discussed below.
The curves labeled ``unitarized'' are the results considering for the 
$T^{\rm{I=1/2}}_{ij}$ amplitudes in Eq.~\eqref{eq:MTunit} the full unitarized amplitudes from  Eq.\eqref{bethes}. The curves labeled  ``BW poles'' are the results using, in Eq.~\eqref{eq:MTunit}, for the final $VP$ amplitude in $I=1/2$ the explicit double Breit-Wigner like shape, Eq.\eqref{eq:MTBW} instead of  Eq.\eqref{bethes}. For this later case, we also show the contribution considering only the lowest mass pole ($A$) or the highest mass pole ($B$) in the Breit-Wigner like amplitude $T^{\rm{I=1/2}}_{ij}$ in Eq.~\eqref{eq:MTunit}. Finally, the curve labeled ``background'' represents the results removing, in Eq.~\eqref{eq:MTunit}, the $T^{\rm{I=1/2}}_{ij}$ amplitude, {\it i.e.}, the $K_1(1270)$ contribution.

First we can clearly appreciate the $K_1(1270)$ resonant effects in the shape of the curves since they differ with respect to the one considering only the background production, (only a) and c) diagrams in Fig.~\ref{fig:TFSI}). Note also that the available phase space does not start at just $m_V+m_P$, because of the finite width of the vector meson. This is more relevant for the $\rho \bar K$ channel.
Next we can compare the result using for the final state interaction in $I=1/2$ the full unitarized amplitudes  with the result using  the explicit double Breit-Wigner like shape, Eq.\eqref{eq:MTBW}. 
The difference is not qualitatively very relevant in the position of the peaks observed. The main physical difference is that the unitarized amplitude in the real axis accounts not only for the possible resonance contributions but it provides the full $VP$ scattering amplitude in $I=1/2$, which then does not necessarily have a Breit-Wigner like shape, or a combination of them. However, in an actual experiment, the resonance would be defined more similarly to a Breit-Wigner (or two if the double pole nature is to be considered) and then an experimental analysis  of the $K_1(1270)$ contribution should obtain something more similar to the ``BW poles'' curve in the figures.
Next, by looking at the contributions considering only the lowest mass pole ($A$) or the highest mass pole ($B$), in the Breit-Wigner like amplitude of Eq.\eqref{eq:MTBW}, we see that in the $D^0\to \pi^+\bar{K}^{*0}\pi^-$ channel the pole $A$ clearly dominates the bump in the mass distribution. This is reminiscent of the large coupling of the lowest mass pole to $K^*\pi$, Table~\ref{tab:coup}. For  the 
$D^0\to \pi^+\rho^0 K^-$  channel
it is worth noting that the available phase space starts close to the $K_1(1270)$ poles and then the allowed phase-space in this region is smaller than for the $\bar K^*\pi$ case. Even though, we can appreciate 
a clear bump closer to the pole $B$. However, the analysis is a bit more involved because of subtle interference effects with the tree level contribution: 
Indeed, in Fig.~\ref{fig:results1}  right we show also the same results as in Fig.~\ref{fig:results1} left but considering only the contribution from the $I=1/2$ VP scattering amplitude, ({\it i.e.} considering only the b) mechanism in Fig.~\ref{fig:TFSI}). This would be the case if one could ideally filter the $K_1(1270)$ contribution. In this case the pole $B$ clearly dominates the  distribution and we can see more clear resonance shapes for the individual pole contributions. 
\begin{figure}[h]
     \centering
     \subfigure[]{\includegraphics[width=.95\linewidth]{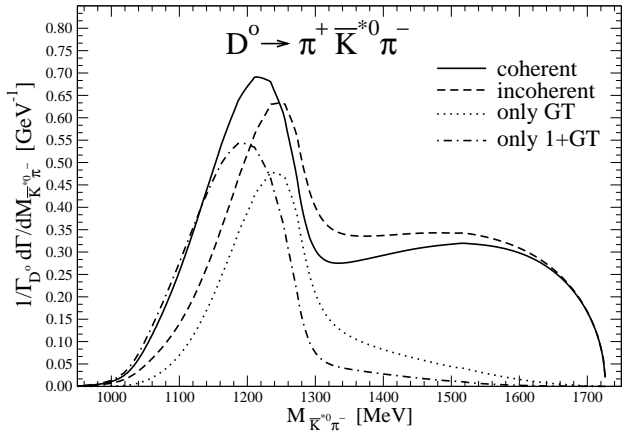}} \\
     \subfigure[]{\includegraphics[width=.95\linewidth]{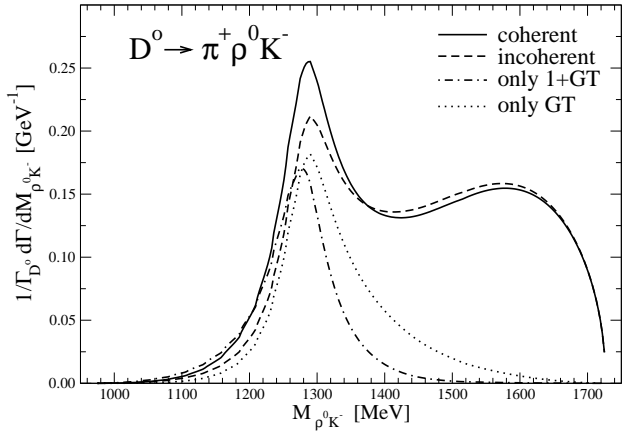}} \\
     \subfigure[]{\includegraphics[width=.95\linewidth]{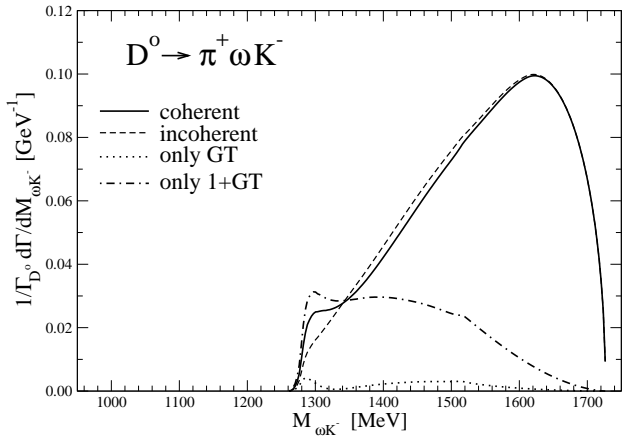}} \\
   \caption{\small{VP invariant mass distribution in the $D^0\to \pi^+ VP$ decay for $D^0\to \pi^+\bar{K}^{*0}\pi^-$,
$D^0\to \pi^+\rho^0 K^-$ and $D^0\to \pi^+ \omega K^-$. Solid line: full results using the unitarized $I=1/2$ amplitude 
with the other contributions added coherently. Dashed line: same but adding the background incoherently.}}
\label{fig:results2}
\end{figure}

In  Fig.~\ref{fig:results2} the  full results (solid line) using the unitarized $I=1/2$ amplitude 
and the other contributions added coherently, as in Eq.~\eqref{eq:MTunit} are compared with the full result but adding the background incoherently,  (dashed line), {\it i.e.}, the sum of curves  ``unitarized'' and ``background'' in  Fig.~\ref{fig:results1} right.
We see that the coherent interference with the background has an important repercussion in the final shape of the distribution.
 Therefore, an experimental analysis should take into account a coherent interference between their $K_1(1270)$ resonance amplitudes and a background term. 
Actually the most important effect is the interference of the unitarized amplitude with the constant term, see first line in 
 Eq.~\eqref{eq:MTunit}. The result considering only the $GT^{\rm{I=1/2}}$ terms in Eq.~\eqref{eq:MTunit}, (curved labeled ``only GT'' in Fig.~\ref{fig:results2}), is also compared to the result adding the constant in front of the summatory in  Eq.~\eqref{eq:MTunit},  (curved labeled ``only 1+GT'' in Fig.~\ref{fig:results2}). We see that the position and shape of the $K_1(1270)$ peaks are clearly distorted.
 This conclusion is also applicable to the other  channels. The important role played by similar interferences with tree level mechanisms in other reactions are also discussed in \cite{Roca:2004uc,sakailiang}.
We can understand why the position and width of the apparent peak differs from the one obtained if one does not consider the interference with the tree level.  If we consider only one channel, for simplicity,
the structure $1+GT$ of Eq.~\eqref{eq:MTunit} is
\begin{align}
1+GT=-\frac{T}{V},
\end{align}
but $V$ has a zero close but below threshold and then $T/V$ tends to accumulate strength in the lowest part of the spectrum and then the resonance appears at lower energies and a bit wider than what is obtained if the interference with the tree level is not considered.

In the $D^0\to \pi^+ \omega  K^-$ channel the available phase space is in principle similar to the $\rho^0 K^-$ channel but the mass distribution does not benefit from the large width of the $\rho$ meson and then the phase space starts more abruptly at $m_\omega+m_K=1278\mev$. However, an accumulation of strength is still visible at the lowest part of the spectrum close the $K_1(1270)$ poles. However, in this observable both poles are responsible for the accumulation of strength because, although the coupling to $\omega K$ is larger for pole $A$, it is further away in energy and then pole $B$ has still a large contribution. Furthermore, there is a strong interference between the amplitudes of both poles. 

From the previous results on the invariant mass distributions we can conclude that the $D^0\to \pi^+\bar K^*\pi$ is the most suited channel to study the lowest mass $K_1(1270)$ pole and the $D^0\to \pi^+ \rho \bar K$ the highest mass one, although both poles should be taken into account and the coherent interference with a nonresonant tree level production must be considered.

\begin{widetext}
\begin{center}
\begin{table}[h!]
\begin{center}
\caption{\small{Different contributions to the branching ratios for $D^0\to \pi^+ VP$ decay. The results in parenthesis are without the background mechanisms. Pole $A$($B$) means that the $B$($A$) pole are removed from the model. (See further explanation in the text). All the results are in $\%$.}}
\begin{tabular}{|c|c|c|c|c|c|}
\hline 
                    &unitarized  & BW poles       & pole $A$   & pole $B$    & background \\ \hline 
$\bar{K}^{*0}\pi^-$ & 23 (8.6)   & 26      (6.6)  & 28 (8.6)   & 14  (0.24)  & 14        \\ \hline 
$\rho^0 K^-$        & 7.9 (2.7)  & 5.8 FIT (0.97) & 5.3 (0.19) & 5.1 (0.57)  & 4.8        \\ \hline 
$\omega K^-$        & 2.8 (0.07) & 2.8 FIT (0.10) &3.0 (0.28)  & 2.8 (0.088) & 2.8        \\
\hline
\end{tabular}
\label{tab:BR}
\end{center}
\end{table}
\end{center}
\end{widetext}

In Table~\ref{tab:BR} we show the different contributions to the integrated branching ratios (in $\%$) for the different  
$D^0\to \pi^+ VP$ channels. The values in the parenthesis are the results removing the background mechanisms.
Note that the results depend on the  $a$ and $b$ coefficients of Eq.~\eqref{eq:aplusb} which, as explained in section~\ref{sec:VPunit},  are obtained from experimental data which has an error of about 30\%.
 Therefore, we can assign to our theoretical results, in Fig.~\ref{fig:results1} and Table~\ref{tab:BR}, a conservative uncertainty of about 50\%. 

In the PDG \cite{pdg} and \cite{Ablikim:2017eqz} there are some experimental branching ratios which could in principle be compared to ours. 
However, all fits to data in the experimental works are based on the existence of a unique $K_1(1270)$ resonance. Should the fits be done assuming explicitly two $K_1$ resonances, with different mass and width and different couplings to ${K}^{*}\pi$ and $\rho K$, the output would certainly be different and, hence, any attempt to compare our results with these experimental ones can only lead to confusion. The value of the present work is that it makes clear predictions for very different shapes with  
${K}^{*}\pi$ and $\rho K$ in the final states. According to this, the optimal way to proceed from an experimental point of view would be, first, to test whether these predictions are correct, and in the positive case proceed to redo fits to data including explicitly two $K_1$ resonances that sum coherently, and background terms in the amplitudes that can interfere with the resonances.

\section{Summary}

We have studied theoretically the suitability of the $D^0\to \pi^+ V P$ decay to check the double pole nature of the $K_1(1270)$ resonance predicted by the chiral unitary approach. Indeed, with the only input of the lowest order chiral perturbation theory  Lagrangians for the $VP$ interaction and the implementation of unitarity in coupled channels, it was obtained in \cite{Roca:2005nm,Geng:2006yb} that the $K_1(1270)$ corresponds actually to two poles located at $(1195-i123)\mev$ and $(1284-i73)\mev$ in the second Riemann sheet of the $VP$ scattering amplitudes, without the need to include the poles as explicit degrees of freedom: they appear naturally from the non-linear dynamics involved in the unitarization procedure. Each pole couples differently to different $VP$ channels and therefore one would expect that reactions weighing differently the mechanisms producing different $VP$ channels would be more influenced by one pole or the other. 
This is the situation in the present case where we have implemented the final state interaction in the elementary $VP$ pair produced in the  $D^0\to \pi^+ V P$ tree level process, where the different channels are related by $SU(3)$. 
We find that the $\bar K^*\pi$ mass distribution in the $D^0\to \pi^+\bar K^*\pi$ channel is clearly dominated by the lowest mass pole and the $D^0\to \pi^+\rho \bar K$ channel by the highest mass pole, although the interference between both poles is relevant. 
This is a consequence of the reactions peaking at different $VP$ invariant masses. We have also discussed the important role played by the coherent interference with the tree level mechanism which should be considered in an experimental analysis in order to properly extract the resonance properties.
Devoted experimental studies of this reaction could help to shed light on the double pole structure of this resonance.

\section{Acknowledgments}

This work is partly supported by the Spanish Ministerio
de Economia y Competitividad and European FEDER funds under Contracts No. FIS2017-84038-C2-1-P B
and No. FIS2017-84038-C2-2-P B, and the Generalitat Valenciana in the program Prometeo II-2014/068, and
the project Severo Ochoa of IFIC, SEV-2014-0398 (EO).
This work is partly supported by the National Natural Science Foundation of China under Grant Nos. 11505158, 11847217.  It is also supported by the Academic Improvement Project of Zhengzhou University. Guan-Ying Wang wishes to acknowledge the support from Zhengzhou University in the program of visiting abroad for Ph.D students.


\begin{thebibliography}{99}


\bibitem{isgurmeson}
  S.~Godfrey and N.~Isgur,
  Phys.\ Rev.\ D {\bf 32}, 189 (1985).

\bibitem{isgurbaryon}
  N.~Isgur and G.~Karl,
  Phys.\ Rev.\ D {\bf 18}, 4187 (1978).
  
\bibitem{capstick}
  S.~Capstick and N.~Isgur,
  Phys.\ Rev.\ D {\bf 34}, 2809 (1986)
  [AIP Conf.\ Proc.\  {\bf 132}, 267 (1985)].
  
\bibitem{roberts}
  S.~Capstick and W.~Roberts,
  Prog.\ Part.\ Nucl.\ Phys.\  {\bf 45}, S241 (2000).
  
\bibitem{vijande}
  J.~Vijande, F.~Fernandez and A.~Valcarce,
  J.\ Phys.\ G {\bf 31}, 481 (2005).
  
\bibitem{sigma}
  J.~R.~Pelaez,
  Phys.\ Rept.\  {\bf 658}, 1 (2016).

\bibitem{ulfreview} 
  F.~K.~Guo, C.~Hanhart, U.~G.~Meissner, Q.~Wang, Q.~Zhao and B.~S.~Zou,
  Rev.\ Mod.\ Phys.\  {\bf 90}, no. 1, 015004 (2018).
  
\bibitem{lutz}
  M.~F.~M.~Lutz and E.~E.~Kolomeitsev,
  Nucl.\ Phys.\ A {\bf 730}, 392 (2004).



\bibitem{Roca:2005nm}
  L.~Roca, E.~Oset and J.~Singh,
  Phys.\ Rev.\ D {\bf 72}, 014002 (2005).
  
\bibitem{chengeng}
  Y.~Zhou, X.~L.~Ren, H.~X.~Chen and L.~S.~Geng,
  Phys.\ Rev.\ D {\bf 90}, no. 1, 014020 (2014).
  
\bibitem{liangsakai}
  W.~H.~Liang, S.~Sakai and E.~Oset,
Phys.\ Rev.\ D {\bf 99},  094020 (2019).
  
\bibitem{dairoca}
  L.~R.~Dai, L.~Roca and E.~Oset,
  Phys.\ Rev.\ D {\bf 99}, no. 9, 096003 (2019).

\bibitem{sakailiang} 
  S.~J.~Jiang, S.~Sakai, W.~H.~Liang and E.~Oset,
  arXiv:1904.08271 [hep-ph].



  
\bibitem{Geng:2006yb}
  L.~S.~Geng, E.~Oset, L.~Roca and J.~A.~Oller,
  Phys.\ Rev.\ D {\bf 75} (2007) 014017.

\bibitem{osetrev}
  E.~Oset {\it et al.},
  Int.\ J.\ Mod.\ Phys.\ E {\bf 25}, 1630001 (2016).
  
  

  
\bibitem{Birse:1996hd}
  M.~C.~Birse,
  Z.\ Phys.\ A {\bf 355}, 231 (1996).

\bibitem{micu}
L. Micu, Nucl. Phys. B {\bf 10}, 521 (1969).


\bibitem{LeYaouanc:1972vsx}
  A.~Le Yaouanc, L.~Oliver, O.~Pene and J.~C.~Raynal,
 Phys.\ Rev.\ D {\bf 8} (1973) 2223.
 
\bibitem{bijker}
  E.~Santopinto and R.~Bijker,   Phys.\ Rev.\ C {\bf 82}, 062202 (2010).



\bibitem{Bramon:1992kr}
  A.~Bramon, A.~Grau and G.~Pancheri,
  Phys.\ Lett.\ B {\bf 283} (1992) 416.



\bibitem{Sakai:2017hpg}
  S.~Sakai, E.~Oset and A.~Ramos,
  Eur.\ Phys.\ J.\ A {\bf 54} (2018) no.1,  10.


  
\bibitem{pdg}
  M.~Tanabashi {\it et al.} [Particle Data Group],
  Phys.\ Rev.\ D {\bf 98} (2018) no.3,  030001.
  doi:10.1103/PhysRevD.98.030001





\bibitem{Roca:2004uc}
  L.~Roca, J.~E.~Palomar, E.~Oset and H.~C.~Chiang,
  Nucl.\ Phys.\ A {\bf 744} (2004) 127.




\bibitem{Ablikim:2017eqz}
  M.~Ablikim {\it et al.} [BESIII Collaboration],
  Phys.\ Rev.\ D {\bf 95} (2017) no.7,  072010.

\end{thebibliography}
\end{document}